\documentclass[aps,pra,floatfix,superscriptaddress,showpacs,twocolumn,amssymb]{revtex4}
\usepackage{amsmath}
\usepackage{mathrsfs}
\usepackage{graphicx}
\usepackage{dcolumn}
\usepackage{bm}
\usepackage{euscript}

\input psfig.sty

\def\bra#1{{\langle#1|}}

\def\cg(#1,#2)(#3,#4)(#5,#6){\bra{#1,#2,#3,#4}#5,#6\rangle}
\def\threej(#1,#2)(#3,#4)(#5,#6){\begin{pmatrix}#1&#3&#5\\#2&#4&#6\end{pmatrix}}
\def\sixj(#1,#2,#3)(#4,#5,#6){\begin{Bmatrix}#1&#2&#3\\#4&#5&#6\end{Bmatrix}}
\def\ninej(#1,#2,#3)(#4,#5,#6)(#7,#8,#9){\begin{Bmatrix}#1&#2&#3\\#4&#5&#6\\#7&#8&#9\end{Bmatrix}}

\begin{document}

\pagenumbering{arabic}

\setcounter{footnote}{0}

\title{Nuclear-spin relaxation of $^{207}$Pb in ferroelectric powders}

\author{L.-S. Bouchard}
 \email{louis.bouchard@gmail.com}
\affiliation{Materials Sciences Division, Lawrence
 Berkeley National Laboratory, Berkeley, California 94720}

\author{A. O. Sushkov}
 \email{alexsushkov@gmail.com}
 \affiliation{Department of Physics, University of California at
Berkeley, Berkeley, California 94720-7300}

\author{D. Budker}
 \email{budker@berkeley.edu}
 \affiliation{Department of Physics, University of California at
Berkeley, Berkeley, California 94720-7300}
 \affiliation{Nuclear Science Division, Lawrence
 Berkeley National Laboratory, Berkeley, California 94720}

\author{J. J. Ford}
\email{Joseph.Ford@pnl.gov} \affiliation{Environmental Molecular
Sciences Laboratory, Pacific North-West National Laboratory,
Richland, Washington 99352}

\author{A. S. Lipton}
\email{AS.Lipton@pnl.gov} \affiliation{Environmental Molecular
Sciences Laboratory, Pacific North-West National Laboratory,
Richland, Washington 99352}

\date{\today}
\begin{abstract}
Motivated by a recent proposal by O. P. Sushkov and co-workers to
search for a P,T-violating Schiff moment of the $^{207}$Pb nucleus
in a ferroelectric solid, we have carried out a high-field nuclear
magnetic resonance study of the longitudinal and transverse spin
relaxation of the lead nuclei from room temperature down to 10 K for
powder samples of lead titanate (PT), lead zirconium titanate (PZT),
and a PT monocrystal. For all powder samples and independently of
temperature, transverse relaxation times were found to be
$T_2\approx 1.5\ $ms, while the longitudinal relaxation times
exhibited a temperature dependence, with $T_1$ of over an hour at
the lowest temperatures, decreasing to $T_1\approx 7\ $s at room
temperature. At high temperatures, the observed behavior is
consistent with a two-phonon Raman process, while in the low
temperature limit, the relaxation appears to be dominated by a
single-phonon (direct) process involving magnetic impurities. This
is the first study of temperature-dependent nuclear-spin relaxation
in PT and PZT ferroelectrics at such low temperatures.  We discuss
the implications of the results for the Schiff-moment search.
\end{abstract}


\maketitle

\section{Introduction}
\label{Section_Intro_older}
The $^{207}$Pb nuclear system (nuclear
spin $I=1/2$; magnetic moment $\mu\approx 0.58\ \mu_N$; isotopic
abundance $\approx 22\%$) in ferroelectric solids has been proposed
for a search for a Schiff moment associated with simultaneous
violation of parity (P) and time-reversal invariance (T) in
fundamental interactions \cite{Muk2005} (see also a discussion of
the sensitivity of such search in Ref. \cite{Bud2006}). The idea is
that, due to the Schiff moment, a ferroelectric sample would acquire
a P,T-odd magnetic polarization along the direction of its electric
polarization.

In a compound free of unpaired electrons, at low enough temperatures
when lattice vibrations freeze out, we have an isolated system of
nuclear spins (in the sense that the spins can freely rotate, but
the nuclei cannot move with respect to the lattice) that interact
among themselves solely via magnetic dipole-dipole interactions. It
was shown in Ref. \cite{Muk2005} that lead nuclei in perovskite
ferroelectric compounds such as lead titanate (PT) experience large
effective electric field,
producing P,T-odd energy shift. The effect is enhanced in
ferroelectric lead compounds due, in part, to the rapid (faster than
$Z^2$) scaling of the effective electric field with the charge of
the nucleus, $Z$, which is also combined with the $\sim Z^{2/3}$
scaling of the Schiff moment, see, e.g., Ref.\ \cite[Sect.
8.5.1]{KhripLamor}. Thus, the ferroelectric compounds proposed for
the Schiff-moment search combine the advantage of a large number of
particles accessible to a solid-state experiment with an enhancement
of the effective electric field similar to that in diatomic
molecules (see, for example, the review \cite{Koz95}).

A crucial requirement for the success of the P,T-violation
experiment is a detailed understanding of the behavior of the
isolated nuclear-spin system, and in particular, its relaxation
properties. The connection of the relaxation and fundamental limits
of sensitivity of condensed-matter P- and T-violation experiments
was discussed in Ref.\ \cite{Bud2006}, and a related question of
magnetic susceptibility was discussed theoretically and modeled
numerically in Ref. \cite{Lud2007}.

In this work, we begin experimental investigation of the lead
nuclear-spin system in the context of this application by studying,
using nuclear magnetic resonance (NMR) techniques, longitudinal
($T_1$) and transverse ($T_2$) relaxation in ferroelectric powders
and a PT monocrystal as a function of temperature in the range
10-293~K. We find that, at high temperatures, these ferroelectrics
undergo magnetic relaxation due to a two-phonon Raman process. At
temperatures much lower than the Debye temperature, relaxation is
dominated by impurities and a single-phonon (direct) process. The
obtained results encourage us to further pursue experimental work
towards a Schiff-moment search in lead-based solid ferroelectrics.

\section{Previous work}
\label{Section_Previous_Studies}
\subsection{Studies of lead NMR}
\label{Subsection_Previous_Studies lead NMR} The body of work on
solid-state NMR of lead prior to 2002 was reviewed in Ref.\
\cite{Dyb2002}. Because of the large atomic number of lead, its
compounds often display large chemical shifts, up to several
thousand ppm. The isotropic and anisotropic parts of the chemical
shift tensor, as well as their temperature dependences have been
shown in Ref.\ \cite{Bus2000} to be sensitive probes of the
crystalline structure. The $^{207}$Pb solid-state NMR study of PT
\cite{Bus2000} was conducted at sample temperatures between -150 and
60$^{\circ}$C. In Ref.\ \cite{Bli2004}, Pb NMR of a single-crystal
ferroelectric lead magnesium niobate (PMN) was studied at
temperatures down to 15 K. The authors investigated the effect of
electric poling on the NMR spectrum. They observed longitudinal
relaxation times for lead in the range 1-10 s.

In some cases, NMR spectroscopy of $^{207}$Pb can be done in
conjunction with the spectroscopy of other nonzero-spin nuclei in
the same compound. For example, Ref.\ \cite{Bal2005} discusses
room-temperature magic-angle-spinning (MAS) NMR of oxygen, titanium
and lead in PbZr$_{1-x}$Ti$_x$O$_3$ (PZT, $0\le x\le 1$) and the
effect of the composition ($x$) on NMR spectra. The present authors
are unaware of any studies of the longitudinal and transverse
relaxation times of $^{207}$Pb in PT and PZT below room temperature
prior to the present work.

\subsection{Studies of relaxation mechanisms}
\label{Subection_Previous_Studies_Relaxation}

Of interest to us here is primarily longitudinal relaxation of
nuclear spins in compounds where all electron angular momenta are
paired (with a possible exception of paramagnetic impurities).
Several mechanisms can lead to such relaxation \cite{Abr61,Abr70}.
These processes typically occur as a result of the modulation of the
crystal
field or ligand field due to lattice vibrations (phonons) and can
be, to some degree, experimentally distinguished by the temperature
and magnetic-field dependence of relaxation.

There are two types of spin-phonon interaction processes: the
``direct process'', in which a spin-flip is caused by the
interaction with a single lattice phonon, and a ``Raman process'',
involving interaction with two phonons. The Raman process has been
shown in Ref. \cite{Veg2006} to be the dominant relaxation mechanism
for $^{207}$Pb nuclei in ionic solids such as lead nitrate above
their Debye temperature T$_D$ (typically 100 K to 200 K). The
nuclear spin-relaxation rate due to this process is proportional to
T$^2$ for T$ > \textrm{T}_D$ (each phonon interaction contributes a
factor of T in this classical regime). However for T$ \ll
\textrm{T}_D$, this process has a T$^7$ dependence ~\cite{Abr61}.
(This can be derived by integrating the Bose-Einstein energy
densities of the two phonons over their frequencies up to the Debye
frequency. Since each energy density contains a factor of frequency
cubed, the resulting dependence is T$^7$.) The relaxation rate for
the Raman process is independent of the magnetic field
\cite{Veg2006}. In Ref. \cite{Bec2006}, longitudinal spin-lattice
relaxation of $^{207}${P}b in polycrystalline samples of
{P}b{M}o{O}$_4$ and {P}b{C}l$_2$ was investigated as a function of
temperature down to the liquid-nitrogen temperature range. A T$^2$
dependence of $1/T_1$ was observed and interpreted as second-order
Raman process. A rather complete discussion of the theory is found
in Ref.\ \cite{Veg2006}, where not only the T$^2$ scaling with
temperature and independence of NMR frequency are reproduced, but
also the magnitude of the relaxation rates is accounted for within
an order of magnitude (for lead molybdate, lead chloride, lead
nitrate, thallium nitrate, thallium nitrite, and thallium
perchlorate). The model explains the relaxation as caused by a Raman
process involving the interactions between nuclear spins and lattice
vibrations via a fluctuating spin-rotation magnetic field.

Unfortunately, more often than not, impurities are responsible for
much of the observed nuclear spin-relaxation. Back in the early days
of NMR, Bloembergen \cite{Blo49} investigated nuclear
spin-relaxation of several (non-Pb) compounds between 300 and 1 K,
and found impurities to be the dominant relaxation agent. Spin
diffusion is essential in affecting relaxation for nuclei that are
far away from an impurity. Dependences of the relaxation rates
$\propto$T and $\propto$T$^{2}$ were observed at low temperature. A
further investigation of nuclear spin-relaxation caused by
paramagnetic impurities was carried out by Blumberg \cite{Blu60}. He
conducted saturation-recovery experiments and found that at short
times, there is a $\sqrt{t}$ dependence of the recovery indicative
of diffusion. At longer times, the recovery is exponential, but the
exponent is proportional to impurity concentration.
%
%
%

\section{Experimental method}
\label{Section_Method}

NMR measurements at the ${}^{207}$Pb frequency were done on a 89-mm
bore, 11.7 T Oxford magnet using a Varian NMR Systems (VNMRS)
spectrometer console running VnmrJ 2.1B software. On this
instrument, the proton frequency was $500.195\ $MHz. The RF carrier
frequency was set to $104.481\ $MHz for $\mathrm{PbTiO_3}$ (PT) and
$104.495\ $MHz for PZT. The samples were lightly packed into a 2
cm-long, 0.5 cm-diameter glass tube and excited using the solenoidal
rf coil of a home-built static probe for cryogenic samples. A
liquid-helium cryostat surrounded the NMR probe and cooled the NMR
tube. The temperature of the sample was monitored using a resistive
sensor with a bridge placed in proximity of the glass tube. The
temporal stability in the temperature control over the range
$10$-$200\ $K was better than $0.1\ $K; however, temperature
fluctuations inside the sample were not directly monitored and could
exhibit higher fluctuations and possibly some internal temperature
gradients. To avoid systematic errors in the relaxation measurements
arising from possible sample temperature fluctuations, inversion
times and echo times were ordered randomly for each experiment and
measurements of $T_1$ and $T_2$ were compared to those from repeated
experiments. We estimate the absolute error in temperature
determination to be less than about 1 K. Pulse widths for a $\pi/2$
rotation ranged between $10$ and $13$ $\mu$s over the temperature
range, with the dependence mainly due to variations in the coil
tuning (Q-factor).

Additional control room-temperature measurements with a lead
titanate sample were carried out with a 7.1-T magnet (proton
frequency 300 MHz) with a Varian CMX Infinityplus spectrometer,
under magic-angle spinning at 11 kHz. Pulse widths for a $\pi/2$
rotation on this instrument was $3$ $\mu$s.
\begin{figure}
\includegraphics[width=7.5 cm]{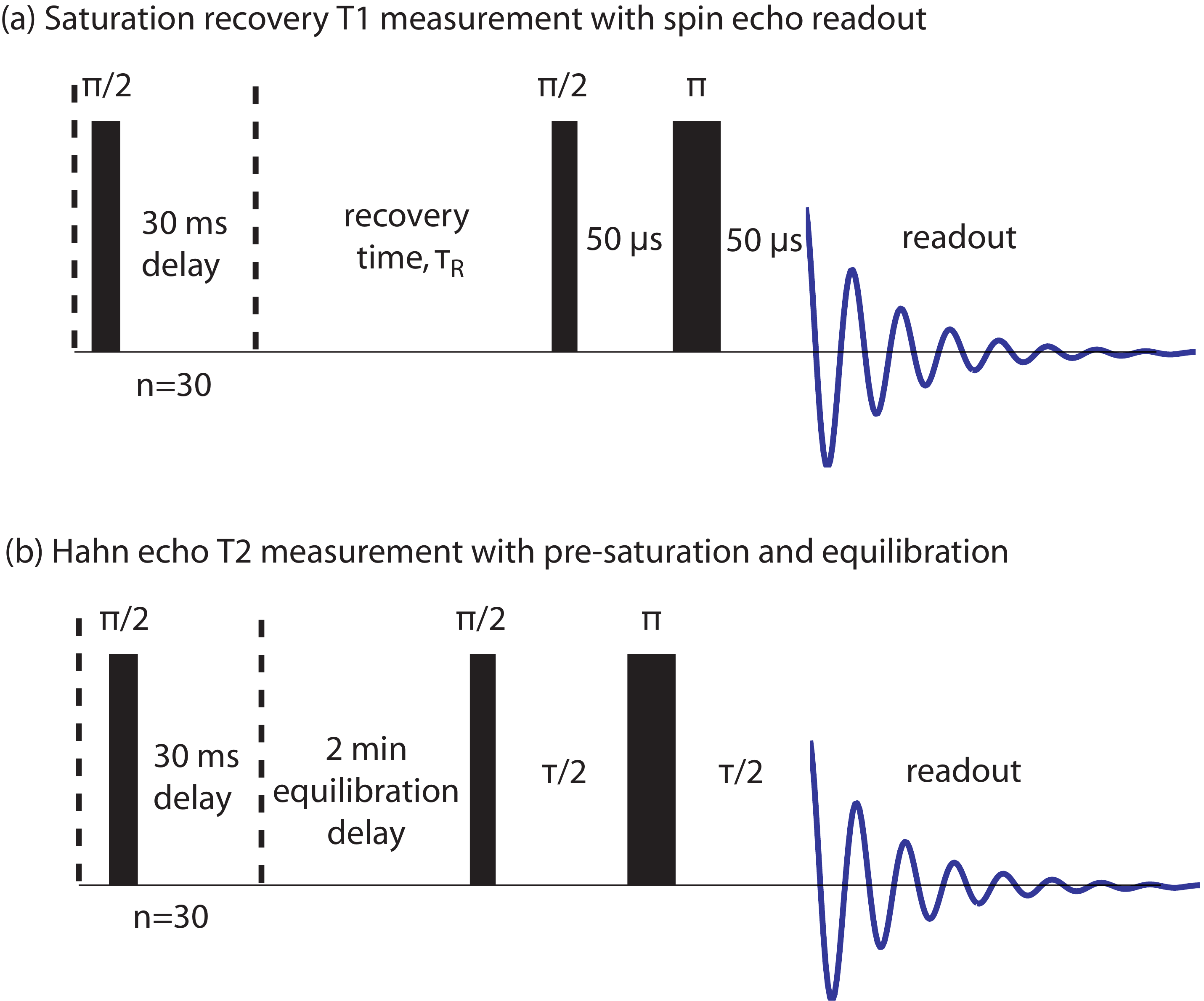}
\caption{\label{fig:pulse_seq} Pulse sequences used for measuring
$T_1$ and $T_2$ relaxation times. In both cases, a train of
 $\pi/2$-pulses was used to saturate the magnetization, followed by a
recovery period. The $T_1$ measurement used a spin echo readout to
avoid coil ringing artifacts and several values of $T_R$ were used.
$T_2$ measurements used a Hahn echo for several values of $\tau$.
The $\pi/2$ pulse widths were in the range of $10-13$ $\mu$s.}
\end{figure}

Transverse ($T_2$) relaxation times were measured on the 11.7-T
system using a $\pi/2-\tau/2-\pi-\tau/2- \{\mbox{acquire}\}$ Hahn
echo [Fig. \ref{fig:pulse_seq} (b)] with echo times ($\tau$) at
$100$ $\mu$s, $200$ $\mu$s, $400$ $\mu$s, $800$ $\mu$s, $1.6$ ms,
$4$ ms, and $8$ ms. The preparation consisted of a saturation pulse
train followed by a $2$-min equilibration period for re-polarization
of the longitudinal magnetization.
At a given temperature, this provided a consistent initial
longitudinal magnetization for each point in a $T_2$ measurement.

\section{Samples}
\label{Section_Samples}

Experiments were performed on four different samples. The PbTiO$_3$
powders were obtained from two different sources: Praxair Specialty
Ceramics (chemical purity 99.9\%, particle size 1 $\mu$m), and from
the group of Prof.~M.~Jansen at Max Plank Institute for Solid State
Research in Stuttgart, Germany (chemical purity unknown). The powder
samples from these two sources are denoted as P1 and P2,
respectively. Measurements were also made on a PbTiO$_3$ crystal
with dimensions 2.3 mm $\times$ 2.8 mm $\times$ 1.3 mm. The fourth
sample was PbZr$_{0.52}$Ti$_{0.48}$O$_3$ powder from Praxair
Specialty Ceramics (chemical purity 99.9\%, particle size 1 $\mu$m).

\section{Results and discussion}
\label{Section_Results}

\subsection{Spin-lattice relaxation}

Longitudinal ($T_1$) relaxation times were measured using a
saturation-recovery sequence shown in Fig.\ \ref{fig:pulse_seq}~(a).
The saturation preparation is accomplished by applying a train of
thirty $\pi/2$ pulses separated  by 30-ms intervals used to
completely depolarize $^{207}$Pb spins in the sensitive volume
despite inhomogeneities in the rf field or inaccuracies in the flip
angle. The saturation was checked by confirming that no signal was
produced in the limit of zero recovery time. Signals were acquired
with 8 to 15 recovery times ($T_R$) logarithmically spaced between
$0\ $ms and $5 \times T_1$. The signals were acquired using a spin
echo with interpulse delay of $50$ $\mu$s to avoid signal
contamination from probe ringing.

All the magnetization-recovery data were fit with two models: a
single exponential {$M(t) = M_0[1-\exp(-t/T_1)]$}, and a stretched
exponential {$M(t) = M_0[1-\exp(-\sqrt{t/T_1})]$}, where $M_0$ is
the equilibrium magnetization, and $t$ is the time elapsed after
saturation. The use of these models is motivated by the fact that
magnetization recovery is well described by a single exponential if
the dominant spin-lattice relaxation mechanism is a two-phonon Raman
process \cite{Veg2006}, or paramagnetic-impurity relaxation with
fast spin diffusion \cite{Sto84}. If, however, the spin-lattice
relaxation is due to paramagnetic impurities with slow
spin-diffusion, then the magnetization recovers along a stretched
exponential curve \cite{Sto84}. It should be noted that the chemical
shift anisotropy in PbTiO$_3$ and PbZr$_{0.52}$Ti$_{0.48}$O$_3$ is
on the order of 2000~ppm, thus, in the applied 11.7-T magnetic
field, the local magnetic field varies by $\simeq200$~G. This is
much greater than the local dipole-dipole interaction field, which
is on the order of 1~G. This does not mean, however, that
spin-diffusion is strongly suppressed. The reason for such a
broadened line is that we are looking at powder.
Each powder grain has a much narrower line,
but, since there is large NMR-shift anisotropy due to the chemical
shift and diamagnetic susceptibility, and the grains are randomly
oriented, the resultant line is wide. Thus, within each powder
crystallite, spin-diffusion can happen, but the spin-flip cannot
``hop'' between the grains.

\begin{figure}[h!]
\includegraphics[width=9cm]{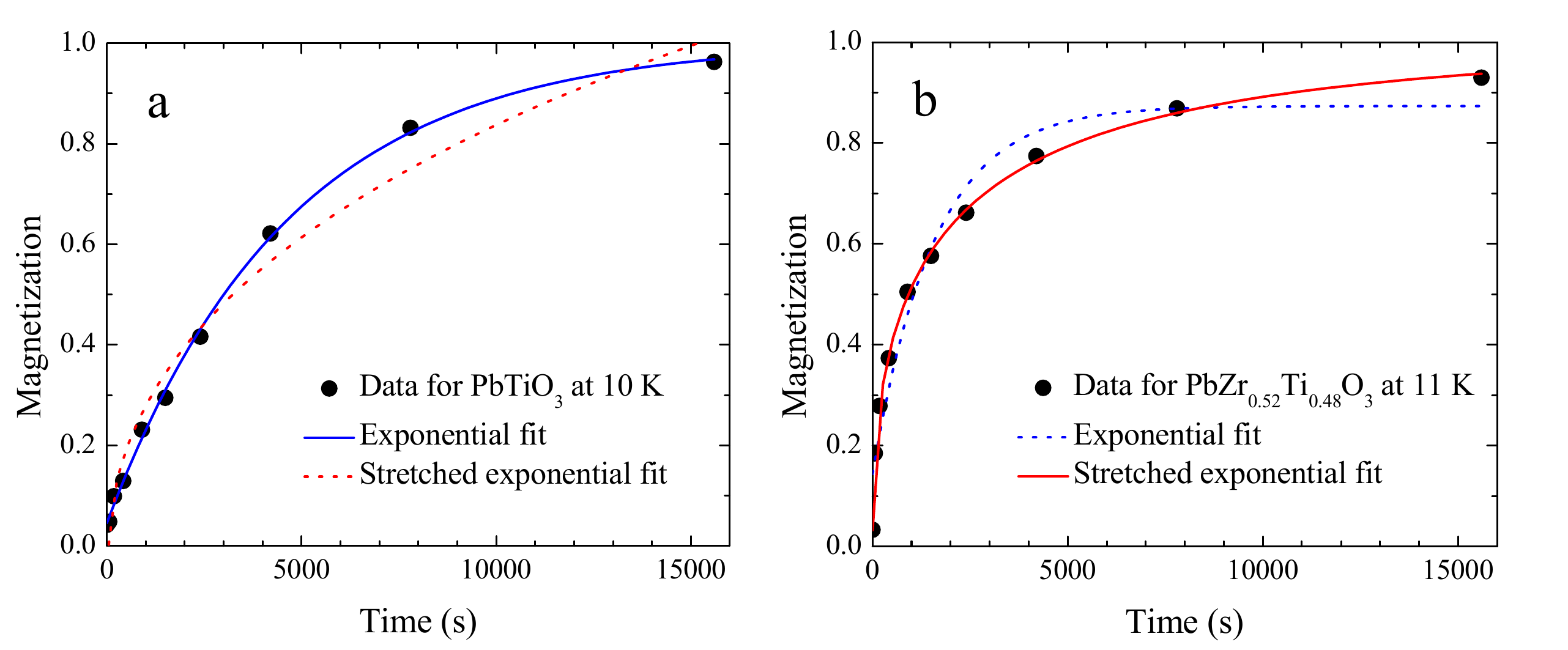}
\caption{\label{fig:MvsTime} Representative magnetization recovery
curves for PbTiO$_3$ (a), and PbZr$_{0.52}$Ti$_{0.48}$O$_3$ (b)
powders. Two fits are shown for each data set: a single exponential
and a stretched exponential. The PT sample is P1 (see text).}
\end{figure}

Magnetization recovery data for PbTiO$_3$ at T$=10$~K are shown in
Fig.~\ref{fig:MvsTime}~(a). The data points are well fit with a
single exponential, a stretched exponential gives a poor fit. This
is the case for all PbTiO$_3$ data at all temperatures. The
situation is different for PbZr$_{0.52}$Ti$_{0.48}$O$_3$. As shown
in Fig.~\ref{fig:MvsTime}~(b), the stretched exponential gives a
much better fit than a single exponential. This is the case for
PbZr$_{0.52}$Ti$_{0.48}$O$_3$ at temperatures below 50~K. For the
data above 50~K, however, single exponential provides a better fit.
This is an indication that spin-lattice relaxation due to
paramagnetic impurities, with suppressed spin-diffusion, may be the
dominant mechanism for PbZr$_{0.52}$Ti$_{0.48}$O$_3$ below 50~K.

\begin{figure}
\includegraphics[width=9cm]{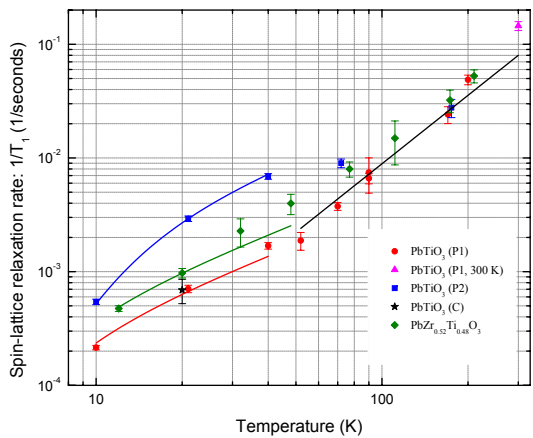}
\caption{\label{fig:T1vsTemp} The experimentally determined
longitudinal relaxation rates vs. temperature. The data were fit to
different functions above and below 50 K (see text).}
\end{figure}

Longitudinal relaxation rates $1/T_1$ extracted from the
magnetization recovery fits as in Fig.~\ref{fig:MvsTime} are plotted
in Fig~\ref{fig:T1vsTemp}.

\subsection{Spin-spin relaxation}

\begin{figure}
\includegraphics[width=7.5 cm]{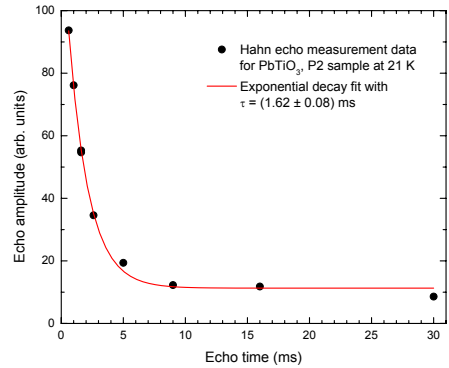}
\caption{\label{fig:T2fit} An example of data and exponential-decay
fit used to extract the $T_2$ values.}
\end{figure}

An example of transverse-relaxation data is shown in Fig.
\ref{fig:T2fit}. Despite of the fact that dipole-dipole interactions
may lead to non-exponential relaxation \cite{Abr61}, within
experiential uncertainties, all our data were consistent with
exponential relaxation, so exponential fits were used to extract the
relaxation times $T_2$.

Analysis of the transverse-relaxation data at different temperatures
results in the value $T_2 = (1.5\pm 0.2)$~ms, for all the PbTiO$_3$
and the PbZr$_{0.52}$Ti$_{0.48}$O$_3$ powder samples. No temperature
variation of $T_2$ is observed within the experimental uncertainty.
(No $T_2$ data were taken for the crystal.)

The measured $T_2$ value is within a factor of two of the
$\approx$0.7-ms decay time estimated for PT in Ref.~\cite{Muk2005}
assuming relaxation due to dipole-dipole interaction. The fact that
the relaxation rate is independent of temperature is also consistent
with relaxation via this mechanism.

\section{Interpretation of $T_1$ results}
\label{Section_Interpretation}

\subsection{High-temperature range}
In the temperature range of 50-293~K, the observed spin-lattice
relaxation rates for all our samples follow the $T^2$ dependence.
This behavior is characteristic of a two-phonon Raman
process~\cite{Abr82,Abr61,Bec2006,Veg2006}, a relaxation mechanism
investigated and explained in terms of modulations in the
spin-rotation interaction by Dybowski and
co-workers~\cite{Bec2006,Veg2006}. Our results, which are the first
measurements below room temperature on PT and PZT perovskite
ferroelectrics, confirm that this relaxation mechanism is, indeed,
the dominant process over the 50-293~K temperature range. We note
that Debye temperatures of ionic crystals are in the range
150-350~K, whereas our results exhibit $T^2$ dependence down to
50~K.

\subsection{Low-temperature range}
At low temperatures (below 50~K) the $T^2$ dependence is no longer a
good fit to the data shown in Fig.~\ref{fig:T1vsTemp}. Another
mechanism dominates spin-lattice relaxation in this regime - this
mechanism is most likely the relaxation due to paramagnetic
impurities in the samples. The $^{207}$Pb nuclear magnetic sublevels
are split by an energy corresponding to about 5~mK in the applied
field of 11.7~T. Phonons which correspond to that same frequency
would normally cause $^{207}$Pb nuclear spin-flips via the direct
process. However, since 5~mK is much smaller than the sample
temperature, the energy density of such phonons is greatly
suppressed. An impurity with uncompensated electron spin, however,
has a much larger magnetic moment. In fact, its magnetic sublevels
are split by about 8~K in the applied magnetic field, which is on
the order of our sample's temperature (at the lowest temperatures
where the data were taken). Therefore such an impurity can interact
with the lattice via the direct phonon process, i.e. absorption or
emission of a single phonon at the Larmor frequency of the
electronic spin $\omega_S$. For a Kramers doublet, the relaxation
rate of the electronic spin is given by the Kramers
relation~\cite{Abr82}:
\begin{equation}
\frac{1}{T_{1e}} = \xi \left( \frac{ \omega_S }{v} \right)^5 \coth
\left( \frac{ \hbar \omega_S} {2k_B \textrm{T} } \right)
\frac{\hbar}{\rho}, \label{Eq:Kramers_Impurity relaxation rate}
\end{equation}
where $v$ is the velocity of sound in the crystal, $\rho$ is the
density and $\xi$ is a dimensionless coefficient that depends on the
structure of the impurity. For non-Kramers sites, there is a
different expression for the relaxation rate, however, the
temperature dependence of $T_{1e}^{-1}$ is the same  as in Eq.
\eqref{Eq:Kramers_Impurity relaxation rate} \cite{Abr70}.

Spin-lattice relaxation of nuclei ($\mathbf{I}$) in the vicinity of
an impurity can be described by a  dipolar coupling $\mathbf{S}
\cdot \mathcal{D} \cdot \mathbf{I}$, where $\mathcal{D}$ is a tensor
describing the magnetic dipolar coupling between the electron and
the nucleus (separated by a distance $\mathbf{r}$), while
$\mathbf{S} \cdot \mathcal{D}$ can be viewed as a random fluctuating
local field produced by the electron and seen by the nucleus. At low
temperatures where motion is frozen, the fluctuations are mainly due
to the spin flipping of the electron, the timescale of which is the
relaxation time $T_{1e}$.

When the power spectrum of these fluctuations contains non-vanishing
values at the Larmor frequency $\omega_I$ of the nucleus, nuclear
relaxation will take place. An argument based on the freezing of
relative motions at low temperatures \cite{Abr82} leads to the
following estimate for the nuclear spin-lattice relaxation rate of
$\mathbf{I}$ due to fluctuations in $\mathbf{S}$:
\begin{equation}
\frac{1}{T_{1n}} (\mathbf{r}) \sim \left( \frac{ B_e }{ B_0 }
\right)^2 \left( \frac{1}{T_{1e}} \right) (1-P_0^2),
\label{Eq:Nuclear relaxation rate}
\end{equation}
where $P_0=\tanh( \hbar \omega_S/2 k_B \textrm{T})$ is the
electronic thermal equilibrium polarization and $B_e= \mu_e S
r^{-3}$ gives the order of magnitude of the local magnetic field
seen by the nucleus due to electron magnetic moment $\mu_e$. A
further level of refinement involves integration of the
spatially-dependent rates ${T_{1n}}^{-1} (\mathbf{r})$ in close
proximity of the impurities ~\cite{Abr82}. Finally, mutual spin
flip-flops of the nuclei cause this relaxation mechanism to spread
its effects throughout the lattice. The (temperature-independent)
rate of spin diffusion and density of impurities determine whether
relaxation is best described by monoexponential or
stretched-exponential behavior \cite{Sto84}.

The low-temperature ($<50$~K) spin-lattice relaxation data were fit
with the function
\begin{equation}
\frac{1}{T_{1n}} = A \coth \left( \frac{ \hbar \omega_S} {2k_B
\textrm{T} }\right)\left[1-\tanh^2\left(\frac{ \hbar \omega_S} {2k_B
\textrm{T} }\right)  \right],
\end{equation}
which is just Eqs.(\ref{Eq:Kramers_Impurity relaxation rate})and
(\ref{Eq:Nuclear relaxation rate}) combined. The results are shown
with curved lines in Fig.~\ref{fig:T1vsTemp}. The fit contained one
free parameter, $A$. The value $\hbar \omega_S/2 k_B = 8$~K was
calculated from the applied magnetic field of 11.7~T (assuming the
impurity g-factor of 2), and fixed for the fits to PbTiO$_3$ P1
powder data and the PbZr$_{0.52}$Ti$_{0.48}$O$_3$ data. It was
found, however, that the PbTiO$_3$ P2 powder data were fit much
better with the value $\hbar \omega_S/2 k_B = 16$~K, i.e. the
paramagnetic impurities in this sample have the magnetic moment of 4
Bohr magnetons. We can conclude that the model described above gives
a satisfactory description of the data below 50~K. The P1 PbTiO$_3$
powder seems to have the smallest concentration of paramagnetic
impurities, while the PbZr$_{0.52}$Ti$_{0.48}$O$_3$ and the P2
PbTiO$_3$ powders have larger impurity concentrations, and the
dominant impurity in PbTiO$_3$ P2 powder has the g-factor of 4.

\subsection{Non-exponential relaxation}

We note that although we fit the relaxation data for PZT to a
stretched exponential for the reasons discussed above, equally good
fits can be obtained by using a biexponential, where the rapid
component is roughly half the amplitude of the slower component.
Possible additional mechanisms which could lead to the
non-exponential behavior include the following. 1) The PZT subunits
PbZrO$_3$ and PbTiO$_3$ may exhibit different spin-rotation
interactions that modulate the relaxation rate $T_1^{-1}$ of the
Raman process~\cite{Veg2006}. 2) The stretched exponential may arise
from a distribution of relaxation times, due to some degree of
anisotropy in the relaxation rate, however, note that such
relaxation behavior is not seen in PT. Unfortunately, the relaxation
data alone are insufficient to determine the source of the deviation
from a single exponential.

\section{Implications for the Schiff-moment experiment}
\label{Section_Implications for Schiff}

The results shown in Fig.~\ref{fig:T1vsTemp} are encouraging for the
ongoing development of the experiment to search for the P- and
T-violating Schiff moment of $^{207}$Pb in ferroelectrics. As
expected, the spin-lattice relaxation time increases dramatically as
temperature is lowered. The presence of paramagnetic impurities
leads to approximately linear temperature dependence of the
relaxation rate below 50~K: $1/T_1\propto$T, which is much slower
than the T$^7$ dependence, expected if the Raman phonon relaxation
mechanism were to dominate. This may prove advantageous for the
design of the Schiff-moment experimental search, which relies on
achieving a high degree of nuclear spin polarization in high
magnetic field and at low temperature, prior to nuclear
demagnetization to reach ultra-low spin temperature~\cite{Bud2006}.
Indeed, with $T_1$ on the order of an hour at lattice temperature of
10~K, such polarization can be established on this time scale, which
is not prohibitively long. After demagnetization, the
spin-temperature stays low for time also on the order of
$T_1\sim$~hour, which is when, for example, the Schiff
moment-induced magnetization can be measured by precision
magnetometry methods.

Two main questions remain to be answered in future investigations:
what is the spin-lattice relaxation time in low magnetic fields and
low spin temperatures, and whether or not the presence of
paramagnetic impurities introduces serious systematics into the
Schiff moment-induced magnetization search. We are currently
developing experiments to address both of these questions.

\section{Conclusion}
\label{Section_Conclusion}

In conclusion, we have presented the first experimental study of
relaxation properties of $^{207}$Pb in PT and PZT below room
temperature. We find that above T$\approx 50\ $K, longitudinal
relaxation rate follows the T$^2$ dependence characteristic of the
two-phonon Raman process. On the other hand, as the temperature is
decreased below T$\approx 50\ $K, the longitudinal relaxation rates
drop slower than $\propto$T$^2$ (as opposed to $\propto$T$^7$
expected for the Raman process), and the relaxation is probably due
to a direct process associated with paramagnetic impurities and
nuclear-spin diffusion. While the longitudinal relaxation times
$T_1$ vary between several seconds and over an hour in the
temperature range between 290 and 10 K, the transverse relaxation
time $T_2$ is found to be $\approx 1.5\ $ms for all temperatures and
all powder samples studied. The obtained results provide an
important input in the design of the experiments to search for
P,T-violating effects in solid ferroelectrics.

\section*{Acknowledgements}
The authors are grateful to Prof.\ A.\ Pines for support,
encouragement and useful discussions, to Prof. E.\ L.\ Hahn for most
useful discussions and inspiration, to Prof. O.\ P.\ Sushkov for
helpful advice, to Prof. R. Ramesh for illuminating discussions of
ferroelectric materials, and to Prof.~M.~Jansen and Mr.
C.~M$\ddot{\textrm{u}}$hle for making and providing us with
ferroelectric-powder samples. This research has been supported by
the Director of the Office of Science of the U.S. Department of
Energy (through the Materials and Nuclear Sciences Divisions of
LBNL). This material is also based in part upon work supported by
the National Science Foundation under Grant No. 0554813. A portion
of the research described in this paper was performed in the
Environmental Molecular Sciences Laboratory, a national scientific
user facility sponsored by the Department of Energy's Office of
Biological and Environmental Research and located at Pacific
Northwest National Laboratory.

\bibliography{Pb207_NMR}
\end{document}